\documentclass{iopconfser}
\usepackage{amsmath,mathtools}
\usepackage{amssymb}
\usepackage{color,xcolor}
\usepackage{siunitx}
\usepackage[greek,english]{babel}
\usepackage{slashed}
\usepackage{appendix}
\begin{document}


\title{Small-scale metric structure and horizons: Probing the nature of gravity 
}

\author{Alessandro Pesci}

\affil{INFN Bologna, Via Irnerio 46, I-40126 Bologna, Italy} 
\vspace{0.1cm}

\email{pesci@bo.infn.it}

\begin{abstract}

A recently developed tool allows for a description of spacetime 
as a manifold with a Lorentz-invariant (lower) limit length built-in.
This is accomplished in terms of geometric quantities depending
on two spacetime events (bitensors) and looking
at the 2-point function of fields on it, 
all this being well suited to embody nonlocality at the small scale. 
What one gets is a metric bitensor  with components singular 
in the coincidence limit of the two events, 
capable to provide a finite distance in the same limit.
We discuss here how this metric structure encompasses
also the case of null separated events,
and describe some results one obtains with the null qmetric
which do have immediate thermodynamic/statistical 
interpretation for horizons. 
One of them is that the area transverse to null geodesics
converging to a base point goes to a finite value in the
coincidence limit (instead of shrinking to 0). 
We comment on the discreteness this seems to imply 
for the area of black hole horizons as well as on possible ensuing 
effects in gravitational waves from binary black hole coalescences.

\end{abstract}

\section{Introduction}

\vspace{0.2cm}

It is quite a general expectation that gravity should exhibit effects
of quantum origin
at very small length scales, possibly
coinciding with the Planck length, or,
also, somehow significantly larger or, perhaps, smaller than that.
It could actually display in principle
quantum features also at much more ordinary scales
--possibly testable in the lab--
in specific circumstances, if it is fundamentally quantum.
The latter is indeed a hot theme of research
with a flourishing of activities, as 
there is no experimental answer to date
as to whether gravity, as we probe it
at macroscopic scales, or at not-exceedingly-small scales,
really needs a quantum description. 

In the very small, however,
a length scale is expected to unavoidably appear
at which some nonlocality,
coming from a `genuine' quantum nature of gravity,
or being somehow induced by the quantum fields of matter
or both, should be present.
What we apparently need anyway,
is that the general-relativistic spacetime
we use in describing gravitational effects,
be endowed somehow, at least at an effective level,
with existence of a limit length;
the universe we inhabit is expected after all to exhibit this feature. 

Aim of the present discussion is to describe a tool
supposed to be used to track these possible/expected quantum
features of gravity at the small scale.
We show in addition how the results one gets this way
point in the direction that existence
of a limit length could leave possibly detectable imprints in the
gravitational wave signals from the merging of
astrophysical compact objects.

\section{Minimum-length metric}

\vspace{0.2cm}

The basic element of quantumness that
the tool we are going to describe considers
is the existence of a distance limit,
foreseen, as hinted to above,
in most full-fledged theories of quantum gravity, at least at
an effective level.
The idea is to have a mathematically and physically well-defined object
which one can use to actually compute distances between
points, having this special property
in the limit of coincidence between them.
This has been successfully accomplished some years ago 
\cite{KotE, KotF, KotI} (the solution being often called minimum-length metric
or quantum metric or qmetric for brief).
Some additions have been put forward
afterwards
for the case of null separations \cite{PesN} and
emphasis will be put on these in the following.
The null case appears quite fit for describing the evolution of horizon,
being the generators of the latter null.

The qmetric then endows spacetime with a limit scale $L$,
the only parameter of the description.
Let us point out first that this is not about
somehow blurring the sources on the scale $L$ and deriving
the modified metric sourced by them.
It is not this. What is maintained instead
is that the existence of a minimum length $L$ affects
geometry itself at the small scale.
What we are talking about is to map the ordinary square geodesic
interval $\sigma^2(x, x')$ between two (time or space separated) points
$x$ and $x'$
(namely $\sigma^2(x, x') = 2 \Omega(x, x')$ with $\Omega(x, x')$ the Synge
function \cite{Synge})
to a new geodesic interval $\sigma^2(x, x') \mapsto S(\sigma^2)$
with $S(\sigma^2) \to \epsilon L^2$ finite ($\epsilon = \pm 1$
for space/time separations; we work in mostly positive signature) 
in the coincidence limit $x \to x'$ 
(with $S(\sigma^2) \approx \sigma^2$ when $|\sigma^2| \gg L^2$,
i.e., when $x$ is far apart from $x'$).
Notice that the prescription just given
corresponds to keeping intact
the causal structure
when going
to the effective metric.
One might envisage perhaps more elaborate possibilities,
and, in particular, signature changes are for sure worth exploring
\cite{KothEucl1, KothEucl2, KothEucl3}
but, in the form we are considering here, we see the qmetric keeps
intact the character of the geodesics (if timelike, or spacelike, ..
or null as we shall see below).
Notice also the choice of introducing an element of quantumness
in the metric description
not searching directly for a quantum metric tensor,
but leveraging instead on something
likely easier to manage (and which still completely describes
the metric properties of the manifold) as the Synge function
\cite{KotE, KotF, KotI}. 

What we have described in the previous paragraph looks like an impossible task.
Indeed, whichever is the metric, the distance has to vanish
in the coincidence; to get a finite limit we would need a metric tensor
singular at that point, which means singular at every point of the manifold.
This indicates that the new description requires quite a strong
departure from ordinary differential geometry.

What we are facing is an expression of the unavoidable nonlocality
of gravity at the
small scale: at the smallest distances things go to be entangled in
kind of finite size blob which represents the physical point there.
This aspect, joined with the singular nature everywhere of the needed
metric structure,
brings to regard bitensors (i.e., tensorial quantities which do depend
on two spacetime points like, first of them, the Synge function itself;
see e.g. \cite{PPV})
as the basic building blocks of such a quantum metric description.

One fixes a point $x'$, which is regarded as base point for distances taken
from it, and considers the other point $x$ as a changing,
field point.
\cite{KotE, KotF} have shown that the fact of requiring
$\sigma^2(x, x') \mapsto S(\sigma^2)$
with $S(\sigma^2) \to \epsilon L^2$ finite in the coincidence limit
$x \to x'$ along the geodesic connecting the two points
amounts to promote the metric tensor
$g_{ab}$ to a metric bitensor $q_{ab}$ which can be written as
\begin{eqnarray}\label{qmetric}
  q_{ab}(x, x') = A \, g_{ab}(x) + \epsilon \, (1/\alpha - A) \, t_a(x) \, t_b(x),
\end{eqnarray}
with $t^a$ the unit tangent vector to the geodesic,
and $\alpha = \alpha(\sigma^2)$, $A = A(\sigma^2)$ two scalar functions
of the geodesic interval, then biscalars themselves.

To fix completely the qmetric, i.e., to provide both $\alpha$ and $A$,
some other condition is needed, additional to the finite limit one.
What does the job is a condition on the propagation of perturbations
in the new, effective metric, that is about the 2-point function
$G(x, x')$ of a field.
To formulate the condition, let us consider first that the new metric gives
rise to a new connection, then to a new covariant derivative,
and then to a new wave operator.
Thinking to the propagation of a massless scalar field,
we consider the d'Alembertian in the new metric.
The condition then consists in requiring that the Green function
of the d'Alembertian in the effective metric in maximally
symmetric spaces be given by the Green function of the d'Alembertian
in the ordinary metric, but evaluated at the shifted geodesic distance $S$
instead of $\sigma^2$,
that is
\begin{eqnarray}\label{ConditionG}
  G(x, x') \mapsto {\widetilde G(x, x')} = G(S(\sigma^2(x, x')),
\end{eqnarray}
where $G$ and $\widetilde G$  are Green functions of the d'Alembertian
operators $\Box$ and $_{x}{\widetilde\Box}_{x'}$ in the ordinary
and in the new metric
$q_{ab}$ respectively \cite{KotI}.
This amounts to require that the wave propagation to some distance
in the new metric is the same as the propagation in the ordinary metric
but at the new square geodesic distance $S$.

This is something very basic to require in any space for which
the Green function depends on the geodesic distance alone,
with no dependence on direction, i.e., on maximally symmetric spaces.
To require the same for generic spaces would be excessive and wrong
since for arbitrary geometries the actual value of $L$ can have
nonnegligible impact on the form of the solution to the
d'Alembert equation;
this will be of lesser and lesser impact the larger the symmetry,
and of no impact at all in case of maximal symmetry.
As a matter of fact, the metric one gets with the addition of
condition (\ref{ConditionG}) is completely fixed: this additional
condition is all we needed, no room for further specifications.
This means that the qmetric we so obtain applies to generic
spacetimes, no particular symmetry required.
Its expression is equation (\ref{qmetric}) with the scalar functions given by
\cite{KotE, KotF, KotI}
\begin{eqnarray}\label{alpha}
  \alpha = \frac{S}{\sigma^2 S'^2}
\end{eqnarray}
and
\begin{eqnarray}\label{A}
  A = \frac{S}{\sigma^2} \,
  \Big(\frac{\Delta}{\Delta_S}\Big)^{\frac{2}{D-1}},
\end{eqnarray}
where $'$ is derivative with respect to the argument ($\sigma^2$ for $S$)
and the expressions are given for $D$-dim spacetime.
$\Delta$ is the van Vleck determinant biscalar (see \cite{PPV})
\begin{eqnarray}
  \Delta(x, x') = - \frac{1}{\sqrt{g(x) g(x')}}
    \det \Big[-\nabla_a^{(x)} \nabla_b^{(x')} \frac{1}{2} \sigma^2(x, x')\Big],
\end{eqnarray}
and
$\Delta_S = \Delta(\widetilde x, x')$ with $\widetilde x$ such that
$\sigma^2(\widetilde x, x') = S$ on the connecting geodesic.

Since $\Delta$, $\Delta_S \to 1$ when $x \to x'$,
we see that the new metric
is singular at any $x'$ in the $x \to x'$ limit as promised,
and this repeats for every $x'$ in spacetime.
Yet, the ``bi'' structure provides a workable expression
describing the metric properties
around any given fixed $x'$.
Notice also that when $|\sigma^2| \gg L^2$, then $S \approx \sigma^2$,
$S' \approx 1$ and $\Delta \approx \Delta_S$, which gives
$\alpha$, $A \approx 1$, and then
$q_{ab}(x, x') \approx g_{ab}(x)$.

\section{Null separations}

\vspace{0.2cm}

The above was for $x$, $x'$ with space and time separations;
what happens if the separation is null?
A first problem we encounter is that, since the square geodesic interval
is identically vanishing along a null ray, what is then the meaning
of a finite distance limit in this case?

\begin{figure}[h]
    \centering
\includegraphics[width=0.5\textwidth]{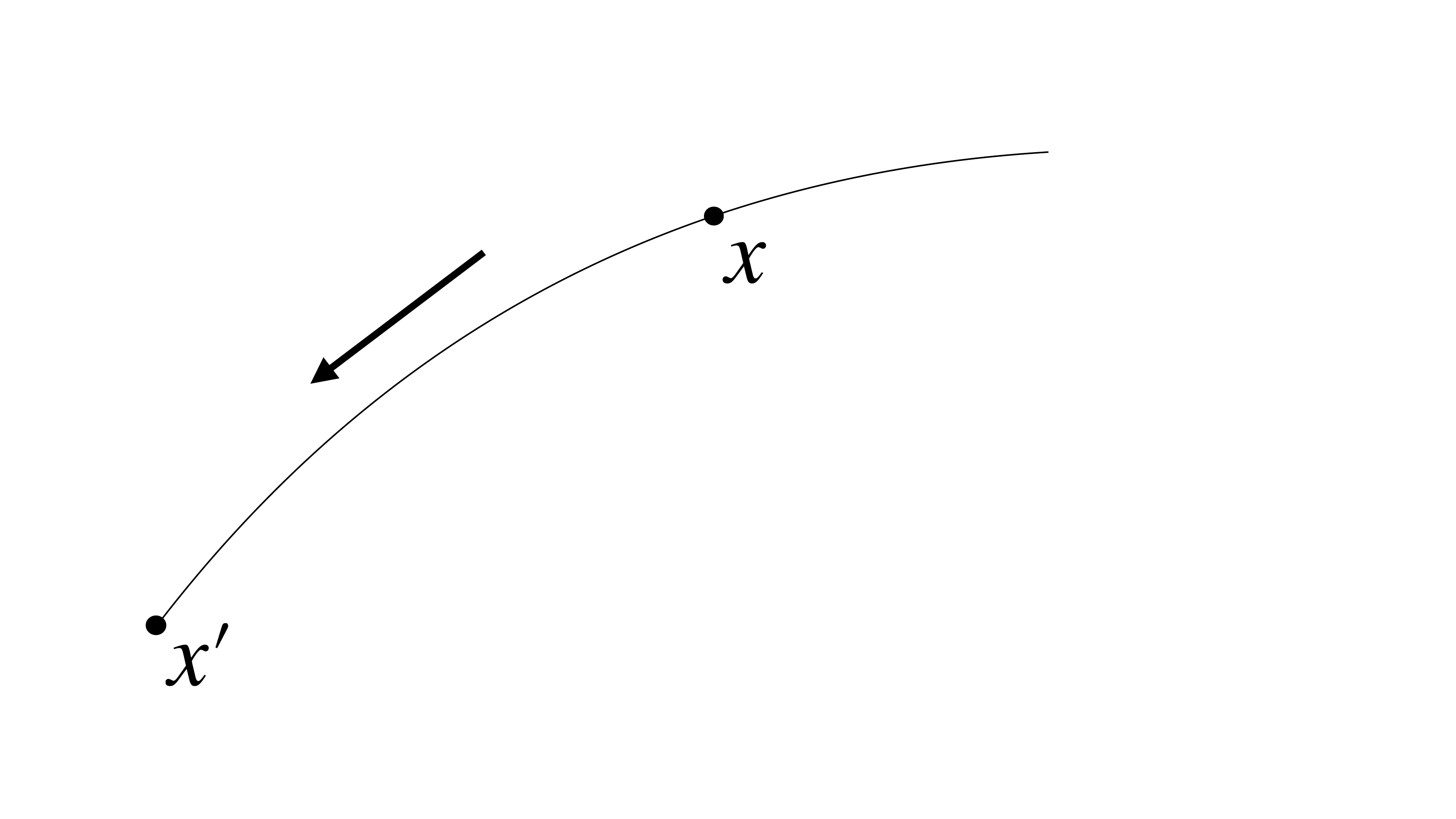}
\vspace{-0.6cm}
\caption{Given a base point $x'$ and a field point $x$,
  the coincidence limit $x\to x'$ is along the
  (assumed unique) connecting geodesic.}
    \label{fig:fig1}
\end{figure}

Let us consider a null geodesic $\gamma$ with affine parameter $\lambda$.
Let us choose $\lambda$ such that $\lambda_{|x'} = 0$;
this means that the coincidence limit
(see Fig.~\ref{fig:fig1}, which applies equally well to null
as well as time or space separations)
is described by $\lambda \to 0$.
Think now of the affine parameter as distance
according to a canonical observer parallel transported
along the geodesic.
This observer at $x'$ will find a finite lower limit $L$ to such distance
when $x \to x'$.
This amounts to require that in the qmetric $\lambda$ gets mapped
to an effective $\widetilde \lambda(\lambda)$,
with $\widetilde\lambda \to L$ when $\lambda \to 0$
(and with $\widetilde\lambda \approx \lambda$ when
$\lambda \gg L$).

Analogously to time and space separations,
we seek a metric bitensor $q_{ab}^{(\gamma)}$ of the form
\begin{eqnarray}\label{null}
  q_{ab}^{(\gamma)}(x, x') =
  A_{(\gamma)} g_{ab}(x) + \bigg(A_{(\gamma)} - \frac{1}{\alpha_{(\gamma)}}\bigg)
  \big(l_a(x) n_b(x) + n_a(x) l_b(x)\big),
\end{eqnarray}  
with $A_{(\gamma)}  = A_{(\gamma)}(\lambda)$
and $\alpha_{(\gamma)} =  \alpha_{(\gamma)}(\lambda)$ two scalar functions
to be determined,
$l^a = (\partial/\partial \lambda)^a$ is tangent to the geodesic
and $n_a$ the usual auxiliary null vector needed to write the metric
transverse to $\gamma$, normalized to $l^a n_a = -1$.

Exactly as it happens for time and space separations
(in which in the effective metric the geodesics remain
affinely parametrized in the new time or space distances),
the geodesic has to remain an affinely parametrized null geodesic
in the new parametrization.
Requiring this,
i.e.,
\begin{eqnarray}
{\widetilde l}^b {\widetilde\nabla_b} {\widetilde l_a} = 0,
\end{eqnarray}
it happens we get straight
\begin{eqnarray}
  \alpha_{(\gamma)} = \frac{C}{d\widetilde\lambda/d\lambda},
\end{eqnarray}
with $C$ a constant \cite{PesN}.
Here ${\widetilde l}^a = \frac{dx^a}{d\widetilde\lambda} =
l^a \frac{d\lambda}{d\widetilde\lambda}$
is the tangent vector to the geodesic in the effective metric
and, for any vector ${\widetilde v^a}$ in effective spacetime,
$ \widetilde\nabla_b {\widetilde v_a} =
\nabla_b {\widetilde v_a}
-\frac{1}{2} q^{cd}
  (-\nabla_d q_{ba} + 2 \nabla_{\left(b\right.} q_{\left.a\right)d}) \, {\widetilde v_c}
$
\cite{KotG}, \cite{PesN}. Note that the indices of objects
are lowered/raised using the ordinary metric when they
are ordinary spacetime indices,
and using the qmetric if they refer to effective spacetime.

\begin{figure}[h]
    \centering
\includegraphics[width=0.5\textwidth]{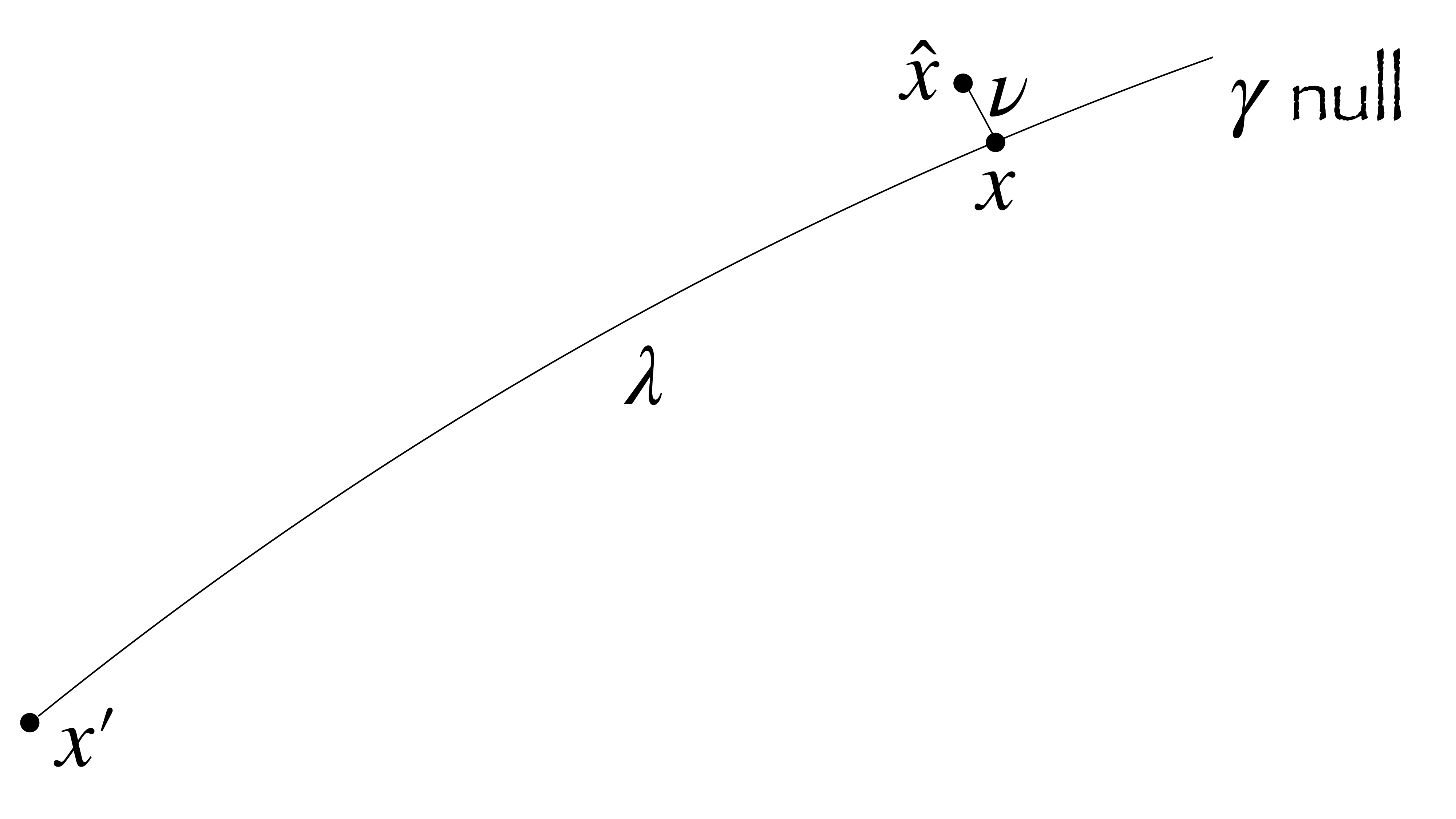}
\vspace{-0.6cm}
\caption{We consider the form of the d'Alembertian at points slightly
  off $\gamma$.}
    \label{fig:fig2}
\end{figure}

Like for time and space separation,
a condition for the d'Alembertian
2-point function for maximally symmetric spacetimes
is also in order.
Here we encounter a second problem specific to null geodesics,
which is the fact that the ordinary 2-point function $G(x, x')$
is diverging for any $x'$, $x$ all along them.
This is tackled by looking at the form that the box operator
acquires not exactly on $\gamma$ but slightly off it (see Fig.~\ref{fig:fig2})
and taking then the limit to $\gamma$ (cf. \cite{VisA}).
For any function $f = f(\sigma^2)$ of the square of the geodesic interval, 
we obtain \cite{PesN}
\begin{eqnarray}
  \Box f = (4 + 2 \lambda \nabla_a l^a) \frac{df}{d\sigma^2}
\end{eqnarray}
with $x$ at separation $\lambda$ from $x'$ on $\gamma$.

The d'Alembertian condition above for maximally symmetric spaces
can then be implemented for null geodesics this way:
${\widetilde G}(x, x') = G(S(\sigma^2(x, x'))$ is solution of
\begin{eqnarray}\label{dAlq}
  (4 + 2 {\widetilde \lambda} {\widetilde \nabla}_a {\widetilde l}^a)
    \frac{d\widetilde G}{dS_{| \widetilde \lambda}} &=&
  (4 + 2 {\widetilde \lambda} {\widetilde \nabla}_a {\widetilde l}^a)
    \frac{d\widetilde G}{d\sigma^2}_{| \lambda = \widetilde \lambda}
  = 0      
\end{eqnarray}
when $G(x, x')$ is solution of
\begin{eqnarray}\label{dAl}
 (4 + 2 \lambda \nabla_a l^a) \frac{dG}{d\sigma^2}_{| \lambda} = 0.
\end{eqnarray}
Using then ${\widetilde\nabla_b} \widetilde l_a$ as provided above
and the expression for $\alpha_{(\gamma)}$ we already have,
Eq. (\ref{dAlq}) is
\begin{eqnarray}
  4 + 2 {\widetilde\lambda} \frac{d\lambda}{d\tilde\lambda}
  \nabla_a l^a_{| \lambda} +
        {\widetilde\lambda} (D-2) \frac{d\lambda}{d{\widetilde\lambda}}
        \frac{d}{d\lambda} \ln A_{(\gamma)} = 0
\end{eqnarray}
for $D$-dim spacetime.

From (\ref{dAl}) at $\widetilde\lambda$,
i.e., $4 + 2 {\widetilde\lambda} \nabla_a l^a_{| \widetilde\lambda} = 0$,
and \cite{VisA}
$\nabla_a l^a_{| \lambda} = \frac{D-2}{\lambda} - \frac{d}{d\lambda} \ln \Delta$,
$\nabla_a l^a_{| \widetilde \lambda} = \frac{D-2}{\widetilde\lambda}
- \frac{d}{d\widetilde\lambda} \ln \Delta_{\widetilde\lambda}$,
we obtain
\begin{eqnarray}
  \frac{d}{d\lambda} \ln
  \bigg[\frac{\lambda^2}{\widetilde\lambda^2}
    \Big(\frac{\Delta_{| \widetilde\lambda}}{\Delta}\Big)^{\frac{2}{D-2}}
    A_{(\gamma)}\bigg] = 0,
\end{eqnarray}
which is
\begin{eqnarray}
  A_{(\gamma)} = C' \frac{\widetilde\lambda^2}{\lambda^2}
  \bigg(\frac{\Delta}{\Delta_{| \widetilde\lambda}}\bigg)^{\frac{2}{D-2}},
\end{eqnarray}
$C' > 0$ constant.

From $q_{ab}^{(\gamma)} \approx g_{ab}$ when $\lambda \gg L$,
we get $C' = 1 = C$.
Then, the final expression we get for null separations
is Eq. (\ref{null}) above with
\begin{eqnarray}\label{alpha_null}
  \alpha_{(\gamma)} = \frac{1}{d\widetilde\lambda/d\lambda}
\end{eqnarray}
and
\begin{eqnarray}\label{A_null}
  A_{(\gamma)} = \frac{\widetilde\lambda^2}{\lambda^2}
  \Big(\frac{\Delta}{\Delta_{| \widetilde\lambda}}\Big)^{\frac{2}{D-2}}
\end{eqnarray}
\cite{PesN, PesW}.

Again, we see that $q_{ab}(x, x')$ is singular at any $x'$ as expected,
yet having a workable expression at any $x \ne x'$
thanks to the ``bi'' structure; this allows to have a metric description
around any given $x'$.
Summing up, what we have seen in this Section
is that the null case,
even if somehow more problematic than the time and space separation
cases, still admits a description within the minimum-length metric.
We go to consider now some applications of the formulas above
to reconstruct some features of the effective spacetime they describe.

\section{Some results}

\vspace{0.2cm}
Our main goal now is to go to see a specific use
of the tool described above when applied to black hole horizons.
But, before that, let us have a look at some general results
with the minimum-length metric, in order to get a feeling
of the characteristics the spacetime acquires in this effective description. 

One most notable thing comes from the consideration
of a quantity, the Ricci scalar $R$, which plays a prominent
role in the classical derivation of field equations
by variational means.
Indeed $R$ acts as Einstein-Hilbert Lagrangian of the gravitational field,
and the field equations one gets have thus a purely geometric meaning,
with apparently no handle to explain all the intriguing connections
with thermodynamics they turn out to have when considered
beside the basic tenets of quantum mechanics (see e.g. \cite{PadN}).

To try to grasp the main features of effective spacetime it is then
quite natural to see first of all what is the fate of the Ricci scalar when
evaluated in the minimum-length metric,
let denote it with $\widetilde R$.  
Well, what we get is an expression which in the coincidence
limit $x \to x'$ becomes
\begin{eqnarray}
  \lim_{x\to x'} \widetilde R(x, x') = \epsilon D R_{ab} t^a t^b + O(L)
\end{eqnarray}
for time and space separations \cite{KotF, KotI},
and
\begin{eqnarray}\label{Rnull}
  \lim_{x\to x'} \widetilde R(x, x') = (D-1) R_{ab} l^a l^b + O(L)
\end{eqnarray}  
for null separations \cite{PesP},
where $O(L)$ cumulatively denotes terms which vanish as $L$, or
as higher powers of $L$, when $L \to 0$. 

\begin{figure}[h]
    \centering
\includegraphics[width=0.5\textwidth]{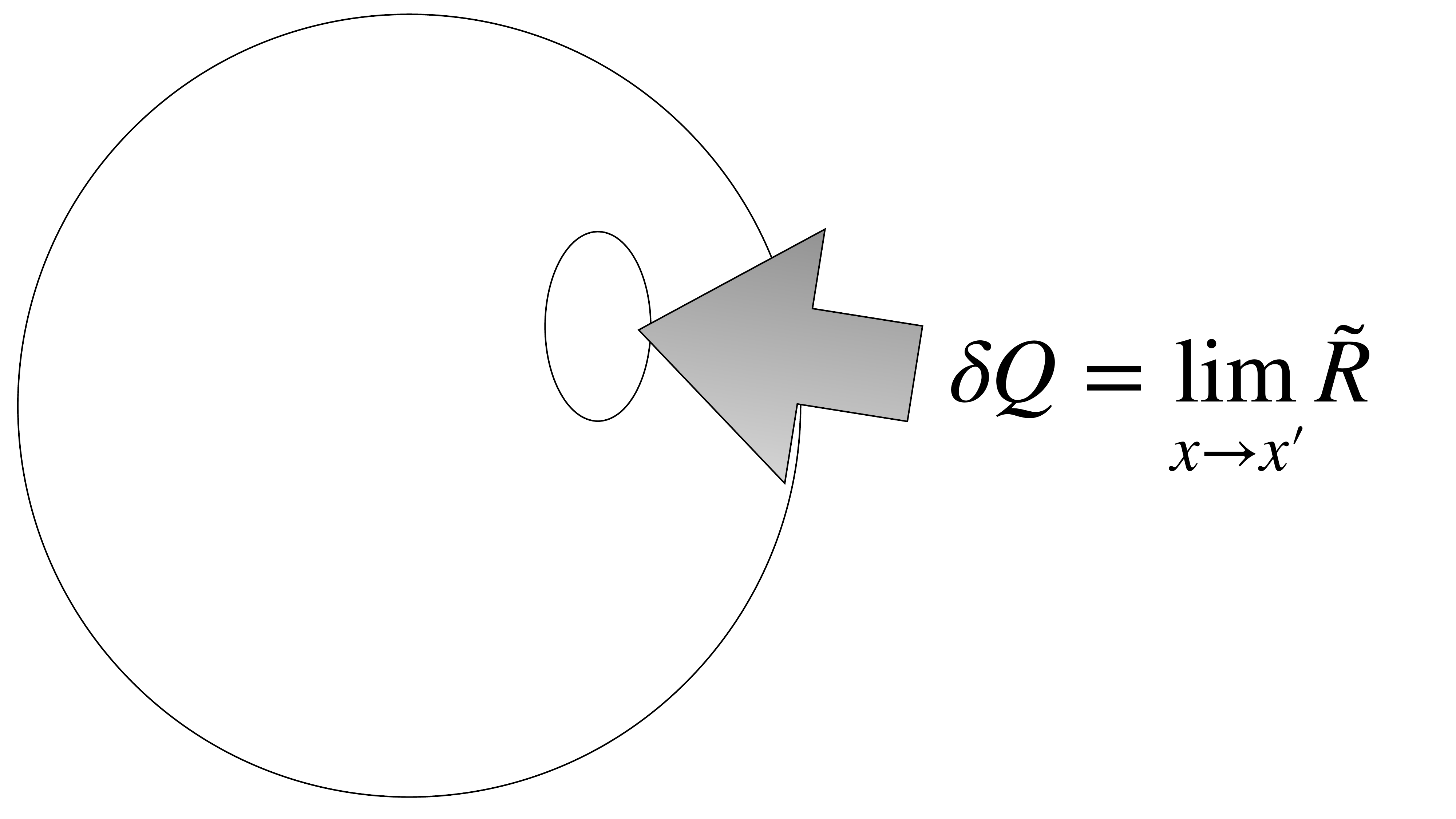}
\vspace{-0.2cm}
\caption{The qmetric Ricci scalar at coincidence turns out to be
proportional to the heat flow of the horizon.}
    \label{fig:fig3}
\end{figure}

There are several things to note about these results;
the one we focus on here is that the leading term in the expression of
$\widetilde R$ at coincidence is proportional to a quantity,
$R_{ab} l^a l^b$ in (\ref{Rnull}),
which is the same quantity that
balances the matter heat flow across the horizon
which has $l^a$ as generator,
when field equations hold true \cite{Jacobson},
see Fig. \ref{fig:fig3}.    
We have thus a quite intriguing consequence:
the existence of a limit length, as implemented by the
minimum-length metric, endows the Ricci scalar with a thermodynamic
meaning,
that is we get,
for free, degrees of freedom to be assigned to gravity and a notion
of entropy density associated to horizons \cite{KotF, KotH}.
The field equations arise from a balance between heat density
variations,
those for gravity and matter \cite{Cha1};
or, explicitly introducing 
a density of gravitational degrees of freedom from the qmetric
\cite{Paddy1, Paddy2},
from an extremization of total entropy density
(computed from matter and gravity degrees of freedom)
\cite{Paddy1, Paddy2, PesK}, something of statistical-mechanical flavor.

The possibility to assign microscopic degrees of freedom to the
gravitational field turns out to be linked to one key property
of the effective metric:
areas transverse to the connecting geodesic go in the coincidence
limit to a finite value instead of shrinking to 0 \cite{Paddy1}.
What happens, more precisely, is that for timelike or spacelike
connecting geodesics the entire $(D-1)$-dim volume transverse
to the geodesic  goes actually to a finite limit at coincidence; 
in case of null geodesics, since they span a $(D-1)$-dim manifold 
it is the $(D-2)$ spacelike volume transverse
to the null geodesic which goes to a finite limit \cite{PesN};
in $D = 4$ we have then that the 2-dim spacelike areas transverse
to the null geodesic go to a finite limit
(it might appear as something intuitively obvious that this should happen
as we have a limit length after all;
this is true
but one should be careful however,
since
a similar calculation
gives for instance 
that the $D$-dim effective volumes go to vanish
in the same limit \cite{Paddy1}
(and, we do not focus on it, this is also something one might expect
in a somehow refined, still intuitive understanding).

This property of the area limit,
which we will consider in more detail 
in a moment, has several consequences.
One is, as mentioned, the possibility to introduce
gravitational degrees of freedom \cite{Paddy1}.
Another is that,
at least when applied to $D$-dim {\it Euclidean} space,
it gives that this space is effectively 2-dim
at the smallest scale \cite{PCK},
resonating with what was put forward first
by \cite{Amb1, Amb2}.
One more consequence comes about when
one considers the evolution of a congruence
of geodesics through Raychaudhuri's equation.
It is clear that the effective existence of a finite limit
area affects the behaviour of geodesics going towards a singularity
or, before that, to a focal point, 
this coming from 
spacetime itself being meant physically as a collection
of coincidence events.
The ensuing modification of the Raychaudhuri equation
has been explored in \cite{KotG, PesPart, CKP}. 

\begin{figure}[h]
    \centering
\includegraphics[width=0.5\textwidth]{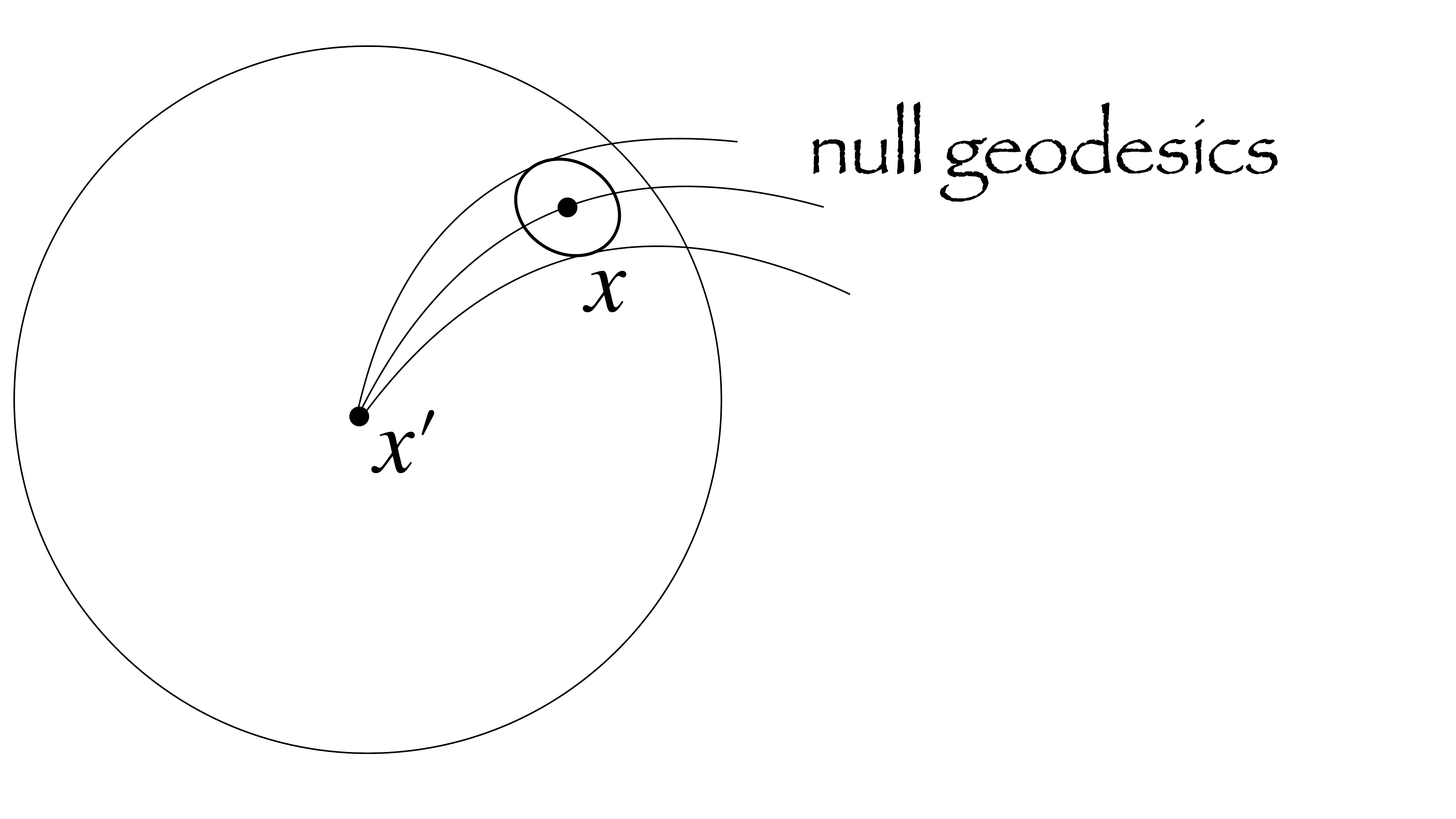}
\vspace{-0.4cm}
\caption{Null geodesics converging to a same base point $x'$.}
    \label{fig:fig4}
\end{figure}

Let us go to describe in some detail
the property of finite limit areas.
We do this because it is something needed
to better see the application we shall describe,
namely the impact of the minimum-length metric
on the evolution of horizon area,
and, precisely because of this, we focus here
on the case of null geodesics.
Let us take a congruence of affinely parametrized
null geodesics emerging from a base point $x'$.
Pick one of them, say the geodesic $\gamma$,
and, at $x$ on it at the value $\lambda$ of the affine parameter,
consider a small (spacelike) surface $S$
of small proper
area $da$ transverse to $\gamma$
on the $(D-2)$-dim equi-geodesic surface at $\lambda$
(see Fig. \ref{fig:fig4}). 
Let imagine then $x$ to go to $x'$, i.e., $\lambda \to 0$,
and the points of $S$ to go to $x'$ with it.

The transverse metric $\widetilde h_{AB}$ in the effective description
($A$, $B = 1,\,  ..., D-2$)
turns out to be given by
\begin{eqnarray}
\widetilde h_{AB} = A_{(\gamma)} h_{AB}
\end{eqnarray}
\cite{KotG, PesN},
where $h_{AB}$ is the ordinary transverse metric.
The effective proper area $d^{D-2} {\widetilde a}(x)$ is then
\begin{eqnarray}
  d^{D-2}{\widetilde a}(x)
  &=&
  \sqrt{\det \widetilde h_{AB}(x) / \det h_{AB}(x)} d^{D-2} a(x) \nonumber \\
  &=&
  \sqrt{\det \widetilde h_{AB}(x) / \det h_{AB}(x)} \lambda^{D-2}
  d\Omega_{(D-2)} \nonumber \\
  &=&
  \widetilde\lambda^{D-2} \frac{\Delta}{\Delta_{| \widetilde\lambda}}
  d\Omega_{(D-2)}
  \to L^{D-2} \frac{1}{\Delta_{| \widetilde\lambda = L}} d\Omega_{D-2}
  \approx L^{D-2} d\Omega_{(D-2)}
\end{eqnarray}
\cite{PesN},
where $d\Omega_{(D-2)}$ is the $(D-2)$-dim element of solid angle.
Here, the second equality comes from the use of (\ref{A_null}),
and the approximation at the end from
$\Delta_{| \widetilde\lambda = L} = 1 + \frac{1}{6} L^2 R_{ab} l^a l^b + ..$
(see e.g. \cite{PPV}).

We see, the effective area in $d\Omega_{(D-2)}$
does not shrink to 0 at coincidence.
Applying this to the case of 4-dim spacetime,
and integrating over the full solid angle,
we obtain the limit value
\begin{eqnarray}\label{limit_area}
  a_0 = 4\pi L^2
\end{eqnarray}
for the area associated at coincidence to point $x'$.
This ought to be regarded as an irreducible area
effectively accompanying any coincidence event,
i.e., any (base) point
in spacetime (cf. \cite{PerriA}).

\section{Use on horizons}

\vspace{0.2cm}

\begin{figure}[h]
    \centering
\includegraphics[width=0.5\textwidth]{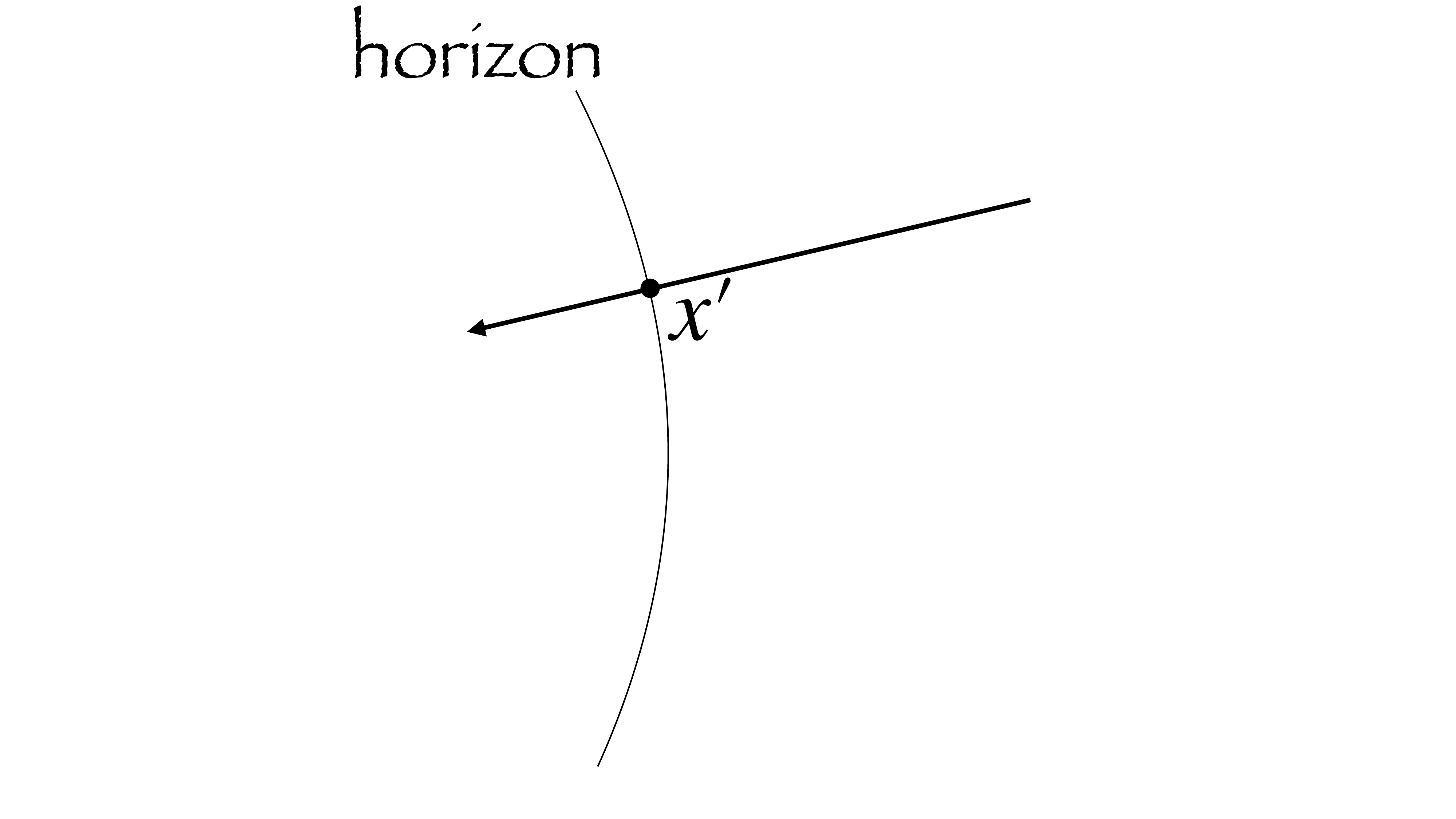}
\vspace{-0.6cm}
\caption{The base point $x'$ is taken as the event of crossing of the horizon
  by some lump of energy.}
    \label{fig:fig5}
\end{figure}

We mention now a specific application of the results above.
A direct consequence of last thing we have considered
is found when it is applied to the null geodesics which build
up the horizon \cite{PerriA}, \cite{KriPer}.
We go to take as point $x'$ the event of crossing
of the horizon by some lump of energy, for instance
that carried by modes of electromagnetic and/or
gravitational radiation (see Fig. \ref{fig:fig5}).
We describe the coincidence event in the frame
of a local observer at $x'$. 

\begin{figure}[h]
    \centering
\includegraphics[width=0.5\textwidth]{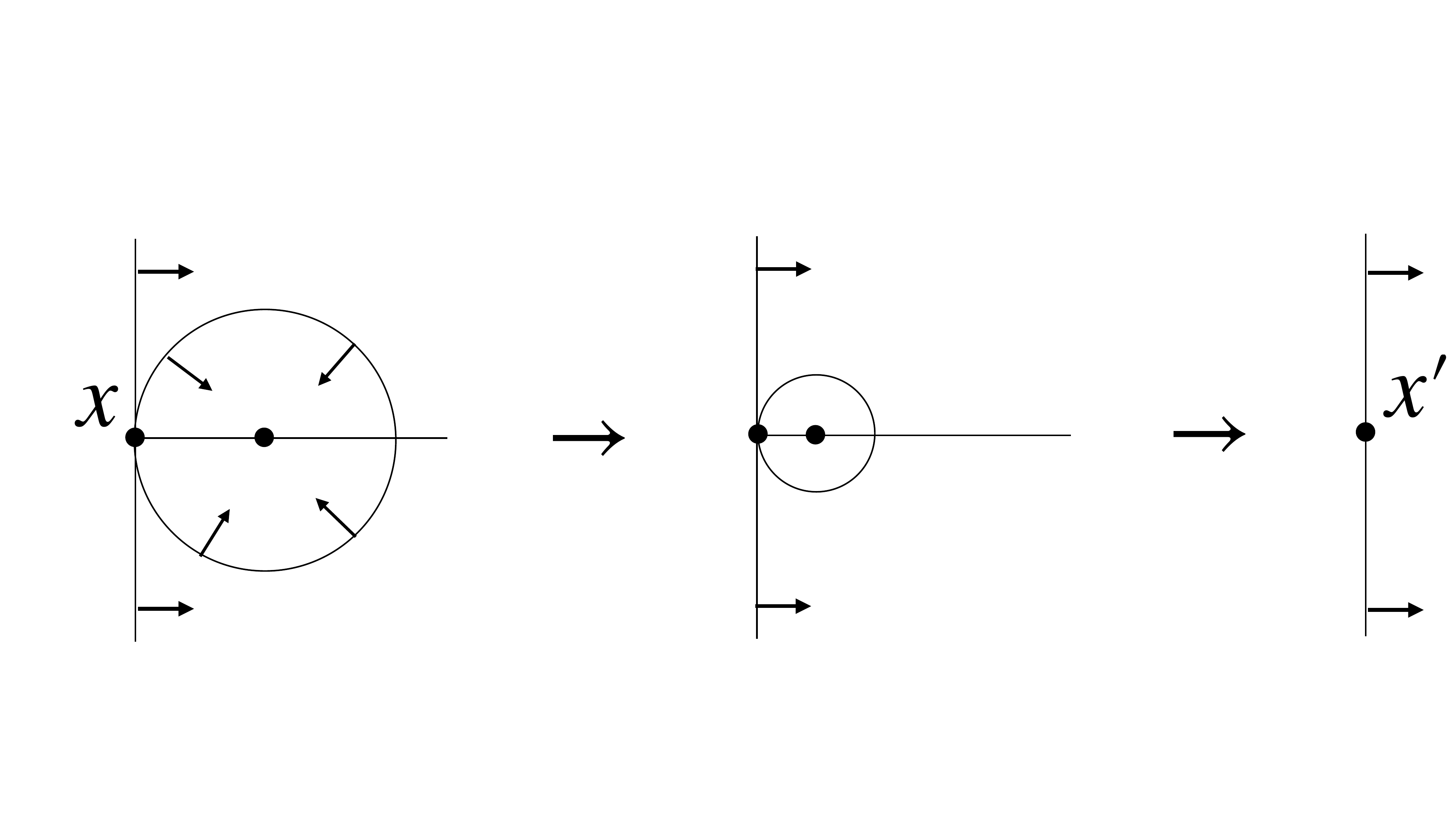}
\vspace{-0.6cm}
\caption{The field point $x$ is on the null geodesic $\gamma$
  which goes to become
  part of the horizon when crossing the lump of energy at event $x'$.
  $x$ is chosen such that it coincides with $x'$ at crossing.
  The figure depicts a $4\pi$ solid angle around the direction of arrival
  along $\gamma$ at $x'$ according to the local observer at $x'$.}
    \label{fig:fig6}
\end{figure}

We take as null geodesic $\gamma$
the geodesic which is just outside the horizon
at start,
and which goes to become part of the horizon
in the process of engulfing of that lump of energy
(see Fig. \ref{fig:fig6}). 

By applying the result (\ref{limit_area}),
we have an irreducible area $a_0$ associated
to the coincidence event $x'$,
sort of effective bump which adds to the area
of the cross-section of the patch of horizon where
the event takes places.
One can then expect that,
after (nearly adiabatic) relaxation, the horizon settles down
to a stationary Kerr, in general, solution
with area ${\cal A'} = {\cal A} + 4\pi L^2$, where $\cal A$ is
horizon area before coincidence.
An increase of the area of the horizon
appears to imply then the overcoming of a minimum area step
$\delta {\cal A}_{\rm min} = 4 \pi L^2$, and this fact
can have in turn huge implications.

A discretization of the horizon area, with steps of the order
of magnitude we obtain here when assuming $L\approx l_p$,
has been as a matter of fact
envisaged by general arguments for quantum black holes
long ago \cite{Bekenstein}
and the effects of this on emission of radiation
have been considered in \cite{BekensteinMukhanov}.
In the absorption,
the engulfing of the lump of energy above means an increase $\delta M$
of the mass of the hole and,
from conservation of energy,
$\delta M$ can not be smaller
than the increase in mass $E_0$ associated to the minimum
area increment $\delta {\cal A}_{\rm min}$,
with $E_0$ depending in general from the mass $M$
and the angular momentum $J$ of the hole.
From $\delta M \ge E_0$,
we have then a minimum energy increase and, in turn,
a minimum angular frequency
$\omega_{\rm min} = E_0/\hbar$ for a mode carrying energy $\hbar \omega$
to be absorbed.
We get then the arising of a reflectivity ${\cal R}\ne 0$ for the black hole;
this
can schematically be modeled by ${\cal R}(\omega) = 1$
when $\omega < \omega_{\rm min}$
and ${\cal R}(\omega) = 0$ when $\omega \ge \omega_{\rm min}$. 
Notice that $\delta {\cal A}_{\rm min}$ and in turn $\omega_{\rm min}$
do depend on the assumed value for the limit length $L$, clearly
both decreasing with the latter. One typically thinks of $L$ as
orders of Planck length $l_p$, but, as mentioned,
scenarios with $L$ significantly
smaller or larger than that should also be considered.

In case of a Schwarzschild black hole we have (in Planck units)
$\delta M = \frac{1}{8\pi} \kappa \, \delta A$,
with $\kappa = \frac{1}{4 M}$ the surface gravity.
This means
$\omega_{\rm min} = E_0 = \frac{\kappa}{2} L^2 = \frac{1}{8 M} L^2$. 
It is interesting to see what this turns out to be in case of a black hole
of say 10 solar masses $M_{\odot}$ and using $L = l_p$.
In terms of frequency we get
$\nu_{\rm min} = \frac{\omega_{\rm min}}{2\pi} =
\frac{1}{16 \pi M} \frac{c^3}{G} = \frac{1}{160\pi} \frac{c^3}{M_\odot G}
\approx 270 \, {\rm Hz}$, well above the frequency cutoff
in sensitivity of current gravitational wave (GW) detectors which is
order of tens of Hz, so that all the frequency range from
the frequency threshold to $\nu_{\rm min}$ is possibly affected by
existence of the minimum area step $\delta A_{\rm min}$.

It is quite amazing that an effect which scales as $l_p^2$
can be in the sensitivity range of currently operating
GW detectors \cite{Agullo}.
Clearly it is not enough that the signal is in the right
frequency range.
Yet, the feasibility of the detection of effects of discretization of
area by units of order $l_p^2$ 
in the inspiral phase or in the relaxation after merger of coalescing
compact objects
has been pointed out \cite{Agullo}
at least for next-generation GW detectors.
What we maintain here is that there is some chance
that effects of this kind could be found
also in current GW detectors,
if only one could pick
the right observable signature.
As for this, observables linked to 
tidal heating during the inspiral phase
seem particularly promising as
the possible effects from $\delta A_{\rm min}$
do accumulate during the process \cite{KriPer}.
Detailed studies of effects of tidal heating on the inspiral
have been reported in \cite{TichyFlanaganPoisson, Alvi, DattaBritoBosePaniHughes, Datta, ChakrabortyDattaSau, KrishnenduChakraborty}.

\section{Conclusions}

\vspace{0.2cm}

We have reported on a tool, the minimum-length effective metric,
able to describe hopefully
in a mathematically well-defined and physically consistent manner 
a spacetime with a minimum length scale $L$ built-in. 
This is obtained by using bitensors as basic building blocks,
prominent among them being the square geodesic interval.
In particular, 
square geodesic intervals between two space separated or time separated
points $x$ and $x'$
go to a finite limit $\pm L^2$, instead of vanishing, in the limit in which
$x$ and $x'$ go to coincide.
It has been shown how this description can also include the case
of null separated events, in spite of being the square geodesic intervals
in this case identically vanishing all along the geodesic.

This last case fits well
for the description of the evolution of the horizons,
as the generators are null. 
When used for this, it foresees the existence of a minimum step
$\delta A_{\rm min} = 4\pi L^2$
for the variation of the horizon area,
and this implies, in turn, the existence of a minimum frequency $\nu_{\rm min}$
(in general depending on mass $M$ and angular momentum $J$ of the black hole) 
for the modes of impinging radiation 
below which they can not be absorbed. 
This induces a reflectivity ${\cal R}(\nu) \ne 0$ for $\nu < \nu_{\rm min}$,
which has generically sizeable overlap with the frequency sensitivity range
of currently operating GW detectors,  
and which has then possible detectable impact on tidal heat absorption
in coalescing black hole binaries \cite{KriPer} as well as
in the ringdown end phase.

The step in area $\delta A_{\rm min}$
we obtain with the minimum-length metric
when we take $L = l_p$
is of the same order of magnitude
as those proposed in literature
since long
for quantum black holes (see \cite{Agullo} and references therein),
coming from treating these
as akin to excited atoms
and recognizing in their area
an adiabatic invariant, thus with area quantization
(the latter being also a prediction of full-fledged theories
of quantum gravity, like in loop quantum gravity \cite{RovelliSmolin},
but with no clear-cut answer concerning the value of the step).

We find quite intriguing that these features, including the value
of the step (in terms of $L$), can be arrived at with
such a simple assumption as the
existence of a limit length and implementing it in the metric
description of spacetime.
This effective metric description can moreover be expected
to be useful
to describe a variety of effects
related to existence of a limit length, more general
than just area quantization of black holes.

\section*{Acknowledgements}
\vspace{0.2cm}

I thank the organizers, and the participants, of
the so interesting DICE2024 meeting
where I gave a talk with the content here reported. 
I thank Sumanta Chakraborty and Max Joseph Fahn
for comments and remarks.
Part of the discussion
is based on unpublished work \cite{KriPer}
done in collaboration with Sumanta Chakraborty,
N. V. Krishnendu and Aldo Perri.
This work is partially supported by INFN grant FLAG.\\

\bibliographystyle{iopart-num}

\bibliography{bib_GravAnom} 

\end{document}